\documentclass[10pt, a4paper]{article}
\usepackage[framemethod=tikz]{mdframed}
\usepackage{lipsum}
\usepackage{dirtytalk}
\usepackage{tcolorbox}
\usepackage[font=footnotesize]{caption}
\newtcolorbox{mybox}[1]{colback=green!6!white,colframe=black!75!black,fonttitle=\bfseries,title=#1}
\newtcolorbox{mybox2}{colback=red!5!white,colframe=red!75!black}

\usepackage{pifont}

\usepackage{soul,xcolor} 
\setstcolor{red} 

\usepackage{xcolor}
\usepackage[hidelinks]{hyperref}
\hypersetup{
   colorlinks,
   linkcolor={blue!50!black},
   citecolor={blue!50!black},
   urlcolor={blue!80!black}
} 

\definecolor{mycolor}{rgb}{0.122, 0.435, 0.698}

\usepackage{titlesec}
\titleformat{\section}
  {\bf\sffamily}
  {\thesection. }
  {5pt}
  {\MakeUppercase}
\renewcommand{\thesection}{\Roman{section}} 

\titleformat{\subsection}
  {\bf\sffamily}
  {\thesubsection. }
  {5pt}{}
  
\renewcommand{\thesubsection}{\Alph{subsection}}

\usepackage[affil-it]{authblk} 
\usepackage{etoolbox}
\usepackage{lmodern}

\usepackage{eurosym}

\usepackage[utf8]{inputenc}
\usepackage{geometry}
 \usepackage[backend=biber,citestyle=numeric,style=phys,biblabel=brackets,sorting=none,url=false,doi=false, natbib]{biblatex}
\bibliography{bibliography}
\DefineBibliographyStrings{english}{andothers={\itshape et\addabbrvspace al\adddot}}

\usepackage{csquotes}
\usepackage[]{authblk} 
\usepackage{etoolbox}
\usepackage{lmodern}

\usepackage{color}
\usepackage{amsmath}
\usepackage{enumerate}
\usepackage{enumitem}
\usepackage{graphicx}
\usepackage{siunitx}
\usepackage{float}
\usepackage[]{parskip} 

\makeatletter
\patchcmd{\@maketitle}{\LARGE \@title}{\fontsize{16}{19.2}\selectfont\@title}{}{}
\makeatother

\usepackage[normalem]{ulem}

\usepackage{subcaption}
\usepackage{color}
\usepackage{graphicx}
\usepackage{dcolumn}
\usepackage{bm}
\usepackage{tikz}
\usetikzlibrary{math}
\usetikzlibrary{shapes.geometric, arrows}
\usetikzlibrary{positioning}
\usetikzlibrary{shapes,arrows}
\usetikzlibrary{intersections}
\usepackage{csvsimple}
\usepackage{changepage}
\usepackage[tbtags]{mathtools}
\usepackage{siunitx}
\usepackage{csquotes}

\usepackage{float}
\usepackage[
	nonumberlist, 				
	acronym,      				
	nomain,						
	nopostdot,					
]{glossaries}
\usepackage{glossary-superragged}
\usepackage[hidelinks]{hyperref}
\usepackage{color, colortbl}

\usepackage{wrapfig}

\usepackage{pgfplots}
\usepackage{tikz}
\usetikzlibrary{arrows}
\usepackage{amsmath}
\pgfplotsset{compat=newest}
\usepgfplotslibrary{fillbetween}
\usetikzlibrary{calc}
\def\centerarc[#1](#2)(#3:#4:#5)
{ \draw[#1] ($(#2)+({#5*cos(#3)},{#5*sin(#3)})$) arc (#3:#4:#5); }

\usepackage{amssymb}

\usepackage{nicefrac}

\usepackage{abstract}

\usepackage{multirow}
\usepackage{tabularx}
\usepackage{booktabs}
\usepackage{array}
\usepackage{longtable}
\newcolumntype{L}[1]{>{\raggedright\let\newline\\\arraybackslash\hspace{0pt}}m{#1}}
\newcolumntype{C}[1]{>{\centering\let\newline\\\arraybackslash\hspace{0pt}}m{#1}}
\newcolumntype{R}[1]{>{\raggedleft\let\newline\\\arraybackslash\hspace{0pt}}m{#1}}

\newacronym{3d}{3D}{three dimensional}
\newacronym{am}{AM}{additive manufacturing}
\newacronym{fdm}{FDM}{fused deposition modeling}
\newacronym{ism}{ISM}{in-space manufacturing}
\newacronym{iss}{ISS}{International Space Station}
\newacronym{fcb}{FCB}{Functional Cargo Block}
\newacronym{dem}{DEM}{discrete element method}
\newacronym{md}{MD}{molecular dynamics}
\newacronym{dc}{DC}{direct-current}
\newacronym[plural=PFCs,firstplural=parabolic flight campaigns (PFCs)]{pfc}{PFC}{Parabolic Flight Campaign}
\newacronym{fft}{FFT}{Fast Fourrier Transform}
\newacronym{cad}{CAD}{Computer Assisted Design}
\newacronym{ptfe}{PTFE}{polytetrafluoroethylene}
\newacronym{ps}{PS}{polystyrene}
\newacronym{nasa}{NASA}{National Aeronautics and Space Administration}
\newacronym{esamm}{ESAMM}{Extended Structure Additive Manufacturing Machine}
\newacronym{amf}{AMF}{Additive Manufacturing Facility}
\newacronym{us}{US}{United States}
\newacronym{usa}{USA}{United States of America}
\newacronym{bmgs}{BMGs}{Bulk Metallic Glasses}
\newacronym{esa}{ESA}{European Space Agency}
\newacronym{si}{SI}{International System of Units, abbreviated from French \textit{Syst\`{e}me International (d'unit\'{e}s)}}
\newacronym{dlr}{DLR}{German Aerospace Center}
\newacronym{liggghts}{LIGGGHTS}{\acrshort{lammps} Improved for General Granular and Granular Heat Transfer Simulations}
\newacronym{lammps}{LAMMPS}{Large-scale Atomic/Molecular Massively Parallel Simulator}
\newacronym{sjkr}{SJKR}{Simplified Johnson-Kendall-Roberts}
\newacronym{ded}{DED}{Directed Energy Deposition}
\newacronym{slm}{SLM}{Selective Laser Melting}
\newacronym{sls}{SLS}{Selective Laser Sintering}
\newacronym{eva}{EVA}{Extra-Vehicular Activity}
\newacronym{sem}{SEM}{Scanning Electron Microscopy}
\newacronym{RPM}{RPM}{Ramdom Positioning Machine}
\newacronym{rpm}{rpm}{revolutions per minute}
\newacronym{rise}{RISE}{Research Internships in Science and Engineering}
\newacronym{daad}{DAAD}{German Academic Exchange Service, abbreviated from German \textit{Deutscher Akademischer Austauschdienst}}
\newacronym{fsm}{FSM}{finite-state machine}
\newacronym{ir}{IR}{infrared}
\newacronym{pcbs}{PCBs}{Printed Circuit Boards}
\newacronym{pcb}{PCB}{Printed Circuit Board}
\newacronym{mcr}{MCR}{Modular Compact Rheometer}
\newacronym{sff}{SFF}{Solid Freeform Fabrication}
\newacronym{uv}{UV}{ultraviolet}
\newacronym{abs}{ABS}{acrylonitrile butadiene styrene}
\newacronym{hpde}{HPDE}{high density polyethylene}
\newacronym{pei}{PEI}{polyetherimide}
\newacronym{bff}{BFF}{BioFabrication Facility}
\newacronym{lens}{LENS}{Laser Engineered Net Shaping}
\newacronym{cnc}{CNC}{Computer Numerical Control}
\newacronym{ebf3}{EBF$^3$}{Electron Beam Free-Form Fabrication}
\newacronym{leo}{LEO}{Low Earth Orbit}
\newacronym{pc}{PC}{polycarbonate}
\newacronym{crissp}{CRISSP}{Customisable Recyclable International Space Station Packaging}
\newacronym{Athena}{Athena}{Advanced Telescope for High-ENergy Astrophysics}
\newacronym{lbm}{LBM}{Laser Beam Melting}
\newacronym{bam}{BAM}{Federal Institute for Materials Research and Testing, abbreviated from German \textit{Bundesanstalt f\"{u}r Materialforschung und-pr\"{u}fung}}
\newacronym{pbf}{PBF}{powder bed fusion}
\newacronym{eb}{EB}{Electron Beam}
\newacronym{2d}{2D}{two dimensional}
\newacronym{4d}{4D}{four dimensional}
\newacronym{ft4}{FT4}{Freeman Technology 4 Powder Rheometer}
\newacronym{dsc}{DSC}{Differential Scanning Calorimetry}
\newacronym{pmma}{PMMA}{polymethylmethacrylate}
\newacronym{1g}{$1g$}{gravity on-ground}
\newacronym{mug}{$\mu g$}{microgravity}
\newacronym{bcm}{BCM}{Box Counting Method}
\newacronym{mct}{MCT}{Mode Coupling Theory}
\newacronym{gmct}{gMCT}{granular Mode Coupling Theory}
\newacronym{itt}{ITT}{Integration Through Transients}
\newacronym{mfc}{MFC}{Mass Flow Controller}
\newacronym{ct}{CT}{computed tomography}
\newacronym{xct}{XCT}{X-ray computed tomography}
\newacronym{cv}{CV}{curriculum vitae}
\newacronym{pi}{PI}{principal investigator}
\newacronym{osp}{OSP}{orthogonal superimposed perturbation}
\newacronym{npi}{NPI}{Network Partnering Initiative}
\newacronym{ecsat}{ECSAT}{European Centre for Space Applications and Telecommunications}
\newacronym{eac}{EAC}{European Astronaut Centre}
\newacronym{estec}{ESTEC}{European Space Research and Technology Centre}
\newacronym{fps}{fps}{frames per second}
\newacronym{pdf}{pdf}{probability density function}
\newacronym{al}{Al}{aluminium}
\newacronym{ss}{\textit{SS}}{\textit{Smooth Surface}}
\newacronym{rs}{\textit{RS}}{\textit{Rough Surface}}
\newacronym{rcp}{rcp}{random close packing}
\newacronym{iop}{IoP UvA}{Institute of Physics of the University of Amsterdam}
\newacronym{mp}{MP}{Institute of Material Physics for Space}
\newacronym{elgra}{ELGRA}{European Low Gravity Research Association}
\newacronym{zarm}{ZARM}{Center of Applied Space Technology and Microgravity}
\newacronym{piv}{PIV}{particle image velocimetry}

\begin{document}

\begin{refsection}

\title{\mbox{Active\hspace{.1em}Cheerios:} 3D-Printed Marangoni-Driven Active Particles at an Interface}
%

\author[1,2]{Jackson K. Wilt \footnote{jacksonwilt@g.harvard.edu, ORCID: 0000-0002-0023-4566}}
\author[1]{Nico Schramma}
\author[1]{Jan-Willem Bottermans}
\author[1]{Maziyar Jalaal\footnote{m.jalaal@uva.nl, ORCID: 0000-0002-5654-8505}}

\affil[1]{Van der Waals-Zeeman Institute, Institute of Physics,\protect\\
University of Amsterdam, Science Park 904, Amsterdam, 1098XH, The Netherlands}
\affil[2]{John A. Paulson School of Engineering and Applied Sciences, Harvard University, Cambridge, MA 02138, USA}

\begingroup
\sffamily
\date{}
\maketitle
\endgroup

\begin{abstract}
Marangoni surfers are simple, cost-effective tabletop experiments that, despite their simplicity, exhibit rich dynamics and collective behaviors driven by physicochemical mechanisms, hydrodynamic interactions, and inertial motion. 
This work introduces self-propelled particles designed and manufactured through 3D printing to move on the air-water interface. We develop particles with tunable motility and controlled particle-particle interactions by leveraging surface tension-mediated forces, such as the Marangoni effect for propulsion and the Cheerios effect for interactions. 
Rapid prototyping through 3D printing facilitates the exploration of a wide design space, enabling precise control over particle shape and function.  We exemplify this by creating translational and chiral particles.
Additionally, we investigate self-assembly in this system and highlight its potential for modular designs where mechanically linked particles with varying characteristics follow outlined trajectories.
This research offers a flexible, low-cost approach to designing active interfacial systems and opens new possibilities for further advancements of adaptive, multifunctional devices.

\end{abstract}


\section{\label{sec:level1}Introduction}
		
Novel propulsion methods for devices moving on fluid interfaces have generated remarkable interest in engineering applications such as chemical delivery, environmental monitoring, and robotic systems~\cite{zhang2011bioinspired,zhang2011bioinspired,yuan2012bio,feldmann2021can,rhee2022surferbot,ho2023capillary,kim2024small,shields_evolution_2017, dong_controlling_2020}. 
While many interfacial devices and living organisms use mechanical forces in fluids to move, several studies have shown that using physicochemical processes at the interface instead proves an effective method to move objects across scales~\cite{bush_walking_2006,deng2022active,pimienta2014self}.
A notable phenomenon in this category is the Marangoni effect, where a surface tension gradient leads to mechanical work~\cite{scriven1960marangoni}.
The Marangoni effect has been known since (at least) the mid 1800s~\cite{thomson1855xlii,marangoni1865sull}, and its utilization by various water-dwelling insects has also been discovered over a century ago~\cite{bush_walking_2006}. 
More recently, various types of synthetic active particles self-propelled by the Marangoni effect and flow have been studied. Examples includes colloids and micro-droplets~\cite{thutupalli2013tuning,dietrich_microscale_2020,hokmabad_emergence_2021,wittmann2021active}, as well as camphor boats~\cite{bourgoin_kolmogorovian_2020,boniface2019self,matsuda_dynamical_2019,hirose_two_2020,matsuda2016acceleration,nishimori2017collective,} and bio-mimetic swimming devices \cite{kwak_marangoni_2021,song2023pen}.
Such ``particles'' break the symmetry with the gradient of surface tension and create a persistent Marangoni self-propulsion lifetime. 

Like many other active systems, precise motility control of Marangoni-driven self-propelled particles is still an open challenge, often due to restrictive fuel properties or design constraints. Previous efforts include changing the shape of the Marangoni surfers~\cite{sur2021effect,renney2013easy}, controlled release of the fuel~\cite{timm2021remotely}, an external electric field~\cite{tsuchitani2020electric} and locally changing the properties of the surface~\cite{tiwari2023capture}. 
Yet, a generic framework to design Marangoni-driven particles with tunable motility and particle-particle interactions offers a broad scope for future work. 
We aim to address this by proposing a simple and frugal system of active particles made by additive manufacturing. Our study is strongly influenced by works typically known for educational purposes, such as studies using household dish soaps and alcohol as fuel. In the case of the colloquially called ``soap boats,'' studies use liquid soap released onto the surface of the water as a powerful surfactant-induced Marangoni effect in classroom demonstrations~\cite{renney_easy_2013}. While soap boats create significant propulsion, the surfactants rapidly saturate the water surface, thus equilibrating the surface tension gradient in very short time spans. An alternative, using common products with rapid diffusion, is ethanol~\cite{ivanov_molecular_2019,burton_biomimicry_2013}. Particles using ethanol solutions as fuel, in one instance, are referred to as \say{cocktail boats}, where an alcohol-driven edible boat was created~\cite{burton_biomimicry_2013}. While both the soap and cocktail boats in the literature are well regarded, they are not designed or investigated for their complex motion. We develop an alcohol (Marangoni)-driven active system following the notion of low cost and accessibility. Meanwhile, we provide an experimental framework to study active particles across a wide range of hydrodynamical regimes, such as inertial active particles, which have received much less attention~\cite{gutierrez2020inertial,antonov2024inertial}.

Our system also utilizes particle-particle interfacial interactions generated from liquid deformation due to the particles. The interface curvature at the particles has shown substantial attraction forces on centimeter-scale objects and surface-dwelling insects~\cite{bush_walking_2006,ko2022small}. This interaction between unconstrained inertial agents, mediated by the surface tension, similar to our system, is best described by the \say{Cheerios effect}, which refers to the clustering behavior observed in floating cereal in a bowl of milk~\cite{vella_cheerios_2005}. The attractive (and repulsive) forces of the Cheerios effect depend highly on the meniscus shape at the solid-fluid interface, governed by material surface properties (hydrophobicity, hydrophilicity) and geometry.

We create our particles with rapid prototyping (3D printing) to efficiently test the space of possible designs. Specifically, the control of fuel deposition is important for the Marangoni effect and the particle geometry for the Cheerios effect.
We design simple centimeter-scale buoy-like particles operated by Marangoni-driven thrust to locomote on an air-water interface. We explore a range of surface tension gradients by varying ethanol concentration~\cite{vazquez_surface_1995} and lowering the surface tension of the fuel mixture, during which we monitor the particle motility and average movement life spans via tracking.
Extending the use of 3D-printing, we explore the concept of chiral particles~\cite{grzybowski2002dynamic,li2024memory,barotta2023bidirectional,scholz_surfactants_2021} and present a Marangoni-driven chiral spinners on the liquid surface. Finally, we show how the present system can benefit modularity, where particles can be linked mechanically to design a geometry with a desirable trajectory.

\section{\label{sec:Sec2}Marangoni-Driven Active Particles Design and Motility}

\begin{figure}[hbt!]
\includegraphics[width=0.95\linewidth]
{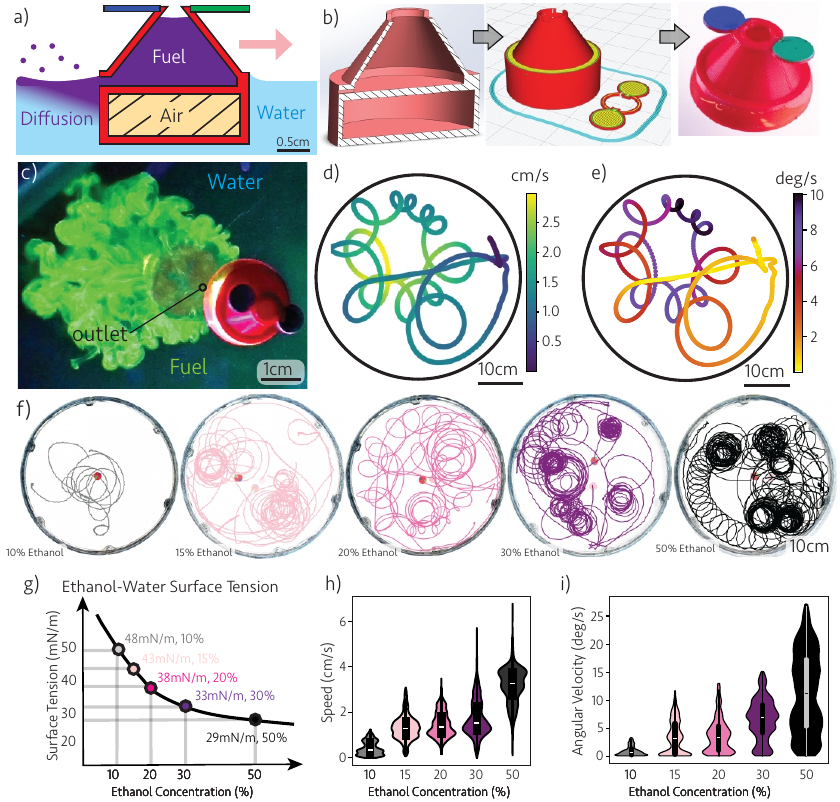}

\caption{Marangoni-driven particle design concept, fabrication, and propulsion mechanics: a) Conceptual design cross-section of the particle. b) Process for 3D design, G-code slicing, and 3D printed final design. c) The particle releases the fluorescent-labeled fuel and self-propels on the surface. d) The trajectory of a single-outlet $50\,\mathrm{\%}$ ethanol-powered particle. The color shows the instantaneous speed of the particle. e) Same as in panel d, but the color shows the instantaneous magnitude of angular velocity. f) Single outlet particle trajectories ($n=3$) with $10\,\mathrm{\%}$, $15\,\mathrm{\%}$  $20\,\mathrm{\%}$, $30\,\mathrm{\%}$, and $50\,\mathrm{\%}$ ethanol fuel concentrations. g) The corresponding surface tension relationships between fuel types (values adopted from~\cite{vazquez_surface_1995}). h) Speed distribution for various ethanol concentrations. i) The angular velocity distribution for various ethanol concentrations.}
 \label{fig:fig1}
\end{figure}

Our active particles are, in principle, a type of self-propelled motor that uses an onboard water-ethanol fuel reservoir and one or multiple small outlets to produce Marangoni-induced forces on a water surface. The surface motors have a buoyant air-filled base, a conical fuel reservoir, one or more fuel release outlets, and a tracking cap (see Fig.~\ref{fig:fig1}a).

Each particle is designed using computer-aided design software (SolidWorks) and exported as a 3D model file to be 3D-printed and assembled, seen in Fig.~\ref{fig:fig1}b, for deployment onto the water surface (see Appendix A for the details of materials used). The tracking caps (blue and green circles) are also 3D printed as an extra module and are attached to the top. 
In single-outlet particles, the blue circle is located on the side of the particle with the fuel outlet, so one can track the orientation of the particle with respect to the outlet position. A simple tracking setup, consisting of a camera located on top of the water basin and particle detection analysis, tracked the position and rotation of the particle through open-source Python packages OpenCV~\cite{opencv_library} and TrackPy~\cite{trackpy}. Particles maneuver in a circular domain (45 cm diameter in Fig.~\ref{fig:fig1}). The black containment rings (also 3D printed) are affixed to the bottom of a much larger water basin ($150\,\mathrm{cm}\,\times\,150\,\mathrm{cm}$). The ring is $3\,\mathrm{mm}$ taller than the water depth of $3\,\mathrm{cm}$ so the dissolution of the fuel is retained if not evaporated; because of this, the ring is removed and replaced after each consecutive test to avoid the accumulation of ethanol however with the largest ring (45 cm) the ring is left in place. 
For optimal repulsion of boundaries and particles, the containment rings are treated with a super-hydrophobic surface coating (NeverWet spray). The combination of a hydrophobic ring and a hydrophilic particle prevents attractive forces between the two, and hence, the rings act like a repulsive wall (see Appendix B). This allows for long measurements of the freely moving particles, avoiding attractive particle-boundary (solid-solid) interactions that could impede their mobility.

The buoyancy of the particles is designed such that the outlets of $500\,\mathrm{\mu m}$ are positioned at the water-air interface (see Fig.~\ref{fig:fig1}a). 
Both direct forcing at the contact line of the particle and Marangoni flow thrust generated at the nozzle play an important role in the particle propulsion (see Appendix C for simple calculations). The fuel (ethanol-water solution) initially exits the tank from this outlet mainly via the capillary actions. Once it arrives at the air-water interface, the fuel changes the local surface tension of the surrounding environment and creates a continuous mass flow rate.
The fuel ejection was visualized by adding $0.1\mathrm{\%}$ fluorescein dye to the fuel mixture and illuminating the basin with a low-power UV-blacklight (see Fig.~\ref{fig:fig1}c for such a visualization for $50\,\mathrm{\%}$ ethanol fuel). A fast flow regime near the outlet is observed, and the fuel plumes show the surface flow generated and surround the fueled motor. Note that the dye concentration can only correlate well with the fuel concentration at the early time of the release, as ethanol evaporates rapidly and leaves the dye in the water basin. We, therefore, use these experiments for demonstrations of initial flow structures only. 


To start, we focus on a single outlet design (see Supplementary Video 1 for an example).
Fig.~\ref{fig:fig1}d shows a typical trajectory and speed of a particle with $50\,\mathrm{\%}$ ethanol solution as fuel.
Driven by the surface tension gradient,
the particles generally show ballistic motion while spinning over their lifetimes. The ballistic motion is expected, given the size and magnitude of the forces in our system. The origin of spinning is most likely akin to the persistently broken symmetry at the point of fuel ejection (also see Appendix D for the angular lag). Nonetheless, the characteristics of this broken symmetry and the associated hydrodynamics are to be fully characterized in future studies. 
The particles' trajectories are constrained and repelled by the hydrophobic containment ring ($45\,\mathrm{cm}$ diameter in the example shown in Fig.~\ref{fig:fig1}d and e), proving the effectiveness of the strategy used to control the particle-boundary interactions. 
The higher speeds of the particles are at the beginning of their lifetimes when higher path curvatures are also observed. As the fuel depletes, the speed of the particle decreases, and the path curvature decreases. Meanwhile, the angular velocity (shown in Fig.~\ref{fig:fig1}e for the same trajectory) also decreases as the particle depletes its fuel, which is closely related to the reorientation associated with the small loops formed from spinning. 
    
We modulate the thrust output from the particle using various ethanol concentrations, which modify the surface tension of the fuel (see Fig.~\ref{fig:fig1}g). We test multiple ethanol-water fuels by mixing deionized water at weight concentrations of $10\,\mathrm{\%}$, $15\,\mathrm{\%}$, $20\,\mathrm{\%}$, $30\,\mathrm{\%}$, and $50\,\mathrm{\%}$ ethanol. Selected trajectories ($n=3$) for these pictures are shown in Fig.~\ref{fig:fig1}f. The ballistic and spiraling characteristics through movement are observed for all fuel constituents. 
The frequency and characteristic size of each spiral, however, correlate with ethanol concentration, with more spiraling occurring at higher concentrations of ethanol. 
As the ethanol concentration increases, we see greater distance traveled with more prominent spiraling characteristics and higher path curvature.
%
The concentration of ethanol controls surface tension (Fig~\ref{fig:fig1}g) and, therefore, the motion characteristics of the Marangoni-driven particle. Panels h and i in Fig.~\ref{fig:fig1} show the distribution of speed and angular velocity for the range of ethanol concentrations used here for particles with single outlets. Both quantities feature a wide distribution of values as particles experience a complex trajectory and also interact with the circular containment wall. However, both the average translational speed ($\sim \mathcal{O} (1\, cm/s)$) and angular velocity ($\sim \mathcal{O} (1-10\, deg/s)$) of the particles increases with increases in ethanol concentration (\emph{i.e.}, increasing the surface tension gradient).

\section{\label{sec:level3} Multi-body Systems: Attraction \& Repulsion in Active Cheerios}

\begin{figure}
\centering
\includegraphics[width=\linewidth]{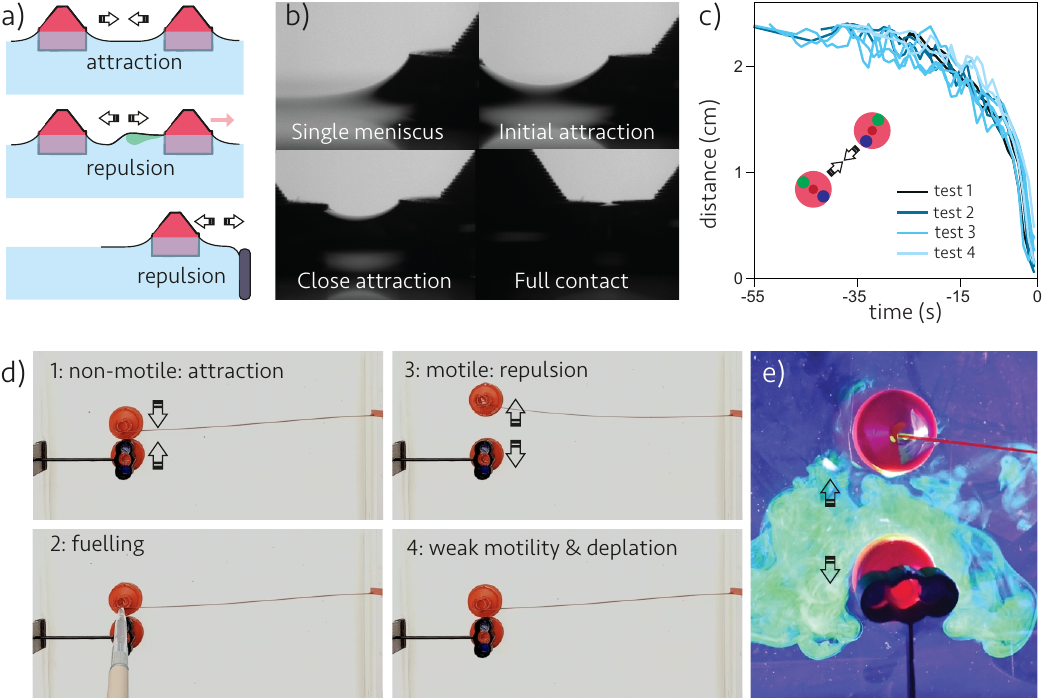}
\caption{Particle interactions on the water surface. a) Hydrophilic particles attract each other on the surface, while an active particle avoids and potentially repels other particles. Meanwhile, a hydrophobic wall repels the particles. b) Side view of two particles attached by deformed interfaces (Cheerios effect). c) The distance between two particles for multiple tests, initially positioned at different locations. d) A demonstration of particle interactions. The top particle is attached to a flexible cantilever, while the bottom particle is fixed in position with a stiff (black) cantilever. Initially, non-motile particles are attached to each other when they are close enough. When the top particle is activated by injecting fuel, the particles repel each other via hydrodynamic forces. Eventually, when the motility weakens, and the fuel is depleted, the Cheerios forces dominate again, and the particles attract each other (see Supplementary Video 3). e) Visualization of the onset of Marangoni flow and repulsion. The bottom particle is unpowered and fixed at its position, while the top particle is attached to the cantilever and fueled with $0.1\,\mathrm{\%}$ fluorescein dye in a $30\,\mathrm{\%}$ ethanol-water mixture.}
\label{fig2:cheerios}
\end{figure}

After tracking single-motor motility, we investigate the attractive and repulsive forces of a two-body system. It is well-known that two bodies positioned on a fluid interface can interact because of the surface deformations they cause~\cite{nicolson1949interaction,vella_cheerios_2005,liu2018capillary,ho2019direct}. Here, the attractive force between the two particles stems from similar hydrophilic particle surfaces. In contrast, the repulsive forces result from the interaction of the particles with hydrophobic walls (see Fig.~\ref{fig2:cheerios}a \& b). 
We first repeat the classic experiments of floating objects on surfaces and find the attraction rate by releasing two confined unfueled particles and monitoring their separation distance. We test different release distances of $4\,\mathrm{cm}$, to $10\,\mathrm{cm}$ (see Fig.~\ref{fig2:cheerios}c) where the Cheerios effect was observed for all these distances. (examples of unpowered attraction seen in Supplementary Video 2) As expected, regardless of the initial position, the distance between the two particles in different tests follows the same rapid attraction and almost identical separation when approaching one another (Fig.~\ref{fig2:cheerios}c). 

\begin{figure}[ht]
\includegraphics[width=\linewidth]
{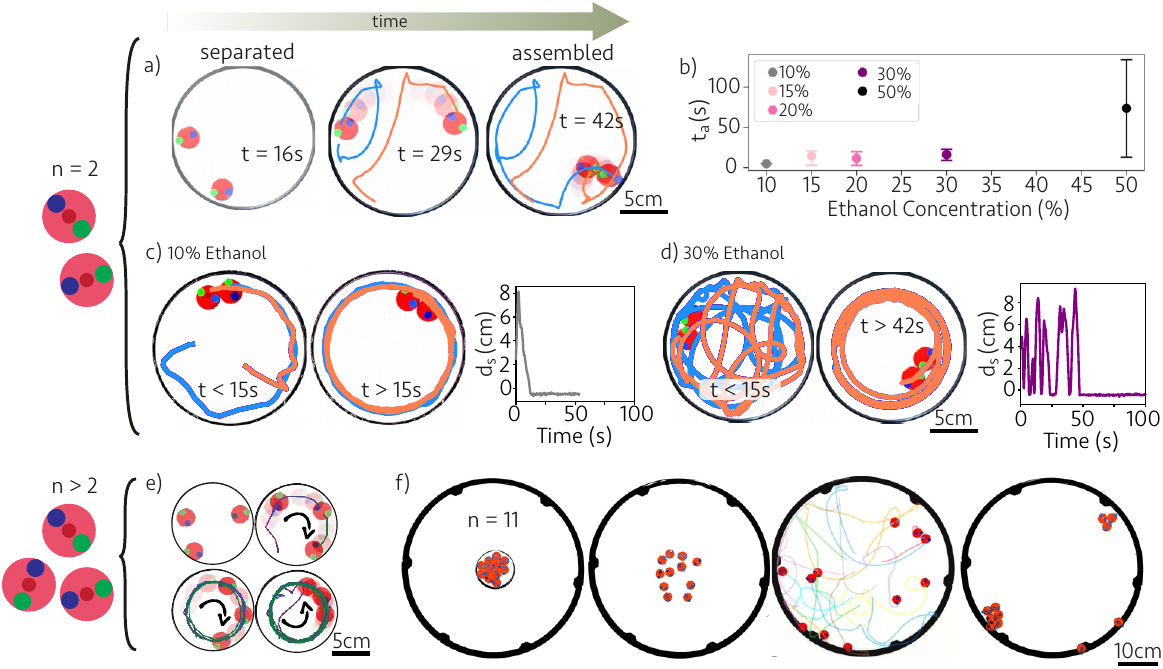}

\caption{Dynamics of multi-body active Cheerios systems. The top and bottom sections indicate the experiments with two or more particles, respectively. a) Example of $30\,\mathrm{\%}$ ethanol fueled particles into its working lifetime at $t=16\,\mathrm{s}$, transit near the containment border at $t=29\,\mathrm{s}$, and its assembled state at $t=42\,\mathrm{s}$. b) Final assembly times for $10\,\mathrm{\%}$, $15\,\mathrm{\%}$, $20\,\mathrm{\%}$, $30\,\mathrm{\%}$, and $50\,\mathrm{\%}$ ethanol concentration fuels. c) Movement up to and after assembly while plotting separation distance ($d_s$) over time of $10\,\mathrm{\%}$ ethanol fuel motors. d) Movement up to and after assembly while plotting separation distance over time of $30\,\mathrm{\%}$ ethanol fuel motors. e) Three active particles in a confinement ring (ratio 6:1), transiting around containment, and aggregation to connected transit around the containment ring (see Supplementary Video 5). f) 11-particle system movement and aggregation using a special dual outlet design with initial fully compact configuration, release, motility, and eventual assembly into multi-particle groups (see Supplementary Video 6).}
\label{fig3:multi}
\end{figure}

The interaction scenarios for active particles are more complex. On the one hand, the Cheerios effect tends to attract the particles, while on the other hand, the active propulsion of each particle causes it to move inertially, primarily avoiding this attraction. As the particle activity decreases, the Cheerios effect becomes dominant, leading to an attraction between particles.
One way to demonstrate this complex interaction between the particles is to constrain their mobility (see Supplementary Video 3). In an experimental setup, we fix one of the particles and allow limited mobility to the other by attaching it to a 3D-printed cantilever beam (see Fig.~\ref{fig2:cheerios}d). Initially, the particles are placed $10\,\mathrm{cm}$ apart and then manually and gradually ($~1\,mm/s$) brought closer together. We approach the static particle until reaching a critical distance, at which the cantilever beam deflects enough for the particles to make contact. This attractive force is based on the Cheerios effect, as explained above, mechanically derived from the contact angle at the water interface and the distance from other interfacial agents \cite{nicolson1949interaction,vella_cheerios_2005}. When fueled, we observe the particle repelling from the adjacent particle, producing an observable deflection. We visualize the onset of Marangoni flow and repulsion with $0.1\,\mathrm{\%}$ fluorescein dye in a $30\,\mathrm{\%}$ ethanol-water mixture (see Fig.~\ref{fig2:cheerios}e). This interaction suggests that the fuel is hydrodynamically interacting with the static particle, likely indicating a force vector that induces deflection. Upon initial fueling, the deflection of the cantilever (and consequently the repulsive force) is at its maximum; this peak in deflection occurs immediately after fueling for all ethanol concentrations. This force diminishes over time, and once the activity weakens and the fuel is exhausted, the active motor rapidly contacts the static motor due to the dominating Cheerios attraction.
It is worth noting that despite our efforts, the resolution and quality of force measurements using the 3D-printed cantilever were insufficient to quantify the propulsion and attraction force values. Alternative methods, such as those used in the work of Ho \emph{et al.}~\cite{ho2019direct}, may be more effective.

\subsection*{\label{sec:level4}Multi-motor Interaction \& self-assembly}
\begin{figure}
\centering
\includegraphics[width=\linewidth]
{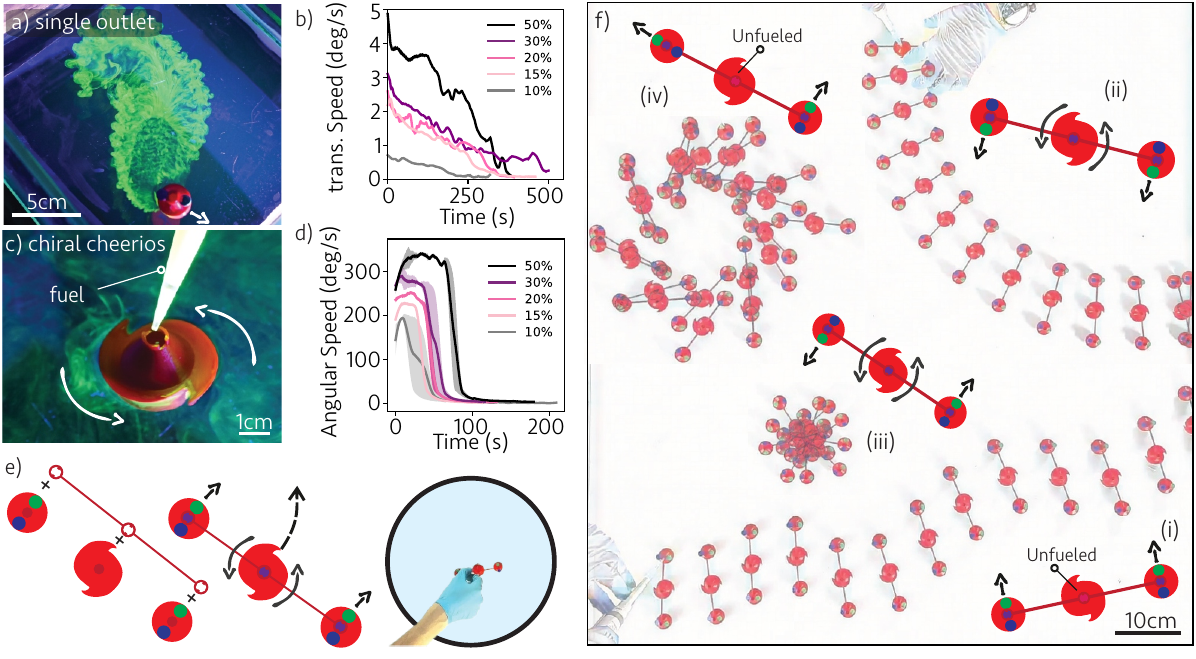}

\caption{Multiparticle construction for engineering propulsion direction. a) Single outlet design with fluorescent fuel. b) Average speed of single outlet particles for different ethanol concentration. c) Chiral particles with fluorescent fuel. d) Angular speed of chiral particles for different ethanol concentration. e) Schematics of modular construction.f) An array of different modular particle assemblies for (i) linear movement, (ii) curved motion, (iii) on-axis spinning, and (iv) off-axis spinning.}
\label{fig:fig4} 
\end{figure}

We proceed to observe and quantify the interaction between multiple powered, confined particles, where the competition between the previously discussed Cheerios effects and the activity of the particles governs the dynamics. The multi-body tests are performed using various hydrophobic ring diameters to investigate separation and assembly properties (see Fig.~\ref{fig3:multi}). The primary size ratio used was $\sim 7:1$ with a confinement ring ($15\,\mathrm{cm}$ diameter) to particles ($2.2\,\mathrm{cm}$ diameter). We release the particles directly next to one another with diametrically opposing outlets to ensure they have the greatest aggregate repulsive thrust. A typical example of a two-body system is shown in Fig.~\ref{fig3:multi}a. (also examples seen in Supplementary Video 4) In the early stages, Marangoni stresses and flows are strong; hence, the activity of the particles dominates, leading to free self-propelled motion. As the particles deplete their fuel, the Cheerios effects become stronger and eventually result in attraction and contact between the particles. The time at which this assembly occurs depends on the ethanol concentration. Generally, the higher the ethanol concentration (and the stronger the motility, as shown in Fig.~\ref{fig:fig1}f \& g), the longer the assembly time, $t_a$.

Low ethanol concentrations do not create enough thrust or inertial momentum to separate the particles for more than several seconds. The particles operating at $15\,\mathrm{\%}$, $20\,\mathrm{\%}$, and $30\,\mathrm{\%}$ ethanol each have marginally longer contact times than those operating at $10\,\mathrm{\%}$ ethanol concentration. Surprisingly, the $50\,\mathrm{\%}$-ethanol-fueled active Cheerios have a much broader range of contact times, from those similar to the order of a few seconds to nearly $120$ seconds. Note that at low motility regimes (see Fig.~\ref{fig3:multi}c for an example), the first contact made by the particles often leads to a permanent assembly. At high motility regimes, the inertial momentum prevents the particles from remaining in contact. Initially, the particles come into contact due to their momentum, carrying them into close proximity and even colliding (where the attractive forces are the strongest). Still, they will separate shortly after the collision due to high inertia. These phenomena are shown in Fig.~\ref{fig3:multi}d, where the particles transiently contact each other up to six times before being dominated by the attractive Cheerios effect when the velocity decreases and the lifetime of the particle reaches its fully assembled state. After the assembly, the two particles enter a circling trajectory at the ring border in almost all cases (see examples in Fig.~\ref{fig3:multi}c \& d). This circling emerges when the particles assemble or repel each other in a self-reinforcing direction of motion around the perimeter, given sufficient fuel reserves. Note that no particular pattern was observed for the relative orientation of the particles upon assembly.

As we increase the number of active Cheerios, the dynamics of collective self-propulsion and self-assembly of particles become more complex. Examples of $3$ and $11$ particles are shown in Fig.~\ref{fig3:multi}e \& f, respectively. The particles in the three-body system (Fig.~\ref{fig3:multi}e) initially expand radially from the center and proceed to transit the perimeter of the containment ring. The direction of circumvention appears to be random, but once established, the particles consistently follow the path in the same direction. (seen in Supplementary Video 5) Finally, we tested an $11$-particle release using a different design with two diametrically opposite-positioned outlet holes that produce opposing thrust, creating an unstable quadrupolar flowfield that rapidly breaks the symmetry and allows the particle to travel transverse to the axis of the outlets (see Fig.~\ref{fig3:multi}e). Two outlets were used to separate the tightly packed particles due to the highly repulsive effects at the beginning of their lifetimes, necessary to overcome Cheerios attraction forces. Note that the two-outlet design has distinctly different motility characteristics from the original single-outlet design but demonstrates aggregation even in a much larger containment ring. Ultimately, when the particles' activities weaken, they assemble in small crystal structures and float in large groups of particles, which are collectively motile until they completely exhaust their fuel (see Supplementary Video 6).

The overall behavior of these systems with a large number of particles depends on various factors, including the design (\emph{e.g.}, number and position of outlets and shape), confinement, ethanol concentration (Marangoni-stress), and clearly the number of particles. A complete parametric study of these factors is beyond the scope of this work. However, we emphasize that the combination of hydrodynamic interactions (flow and interfacial effects) and the stochastic characteristics of the multi-body active Cheerios makes it a significantly rich, albeit frugal, experimental system to explore collective active motions, with many details yet to be discovered. Furthermore, the system's flexibility allows for the design of active systems, a goal we will pursue in the next and final sections.

\section{\label{sec:level5}Chiral particles and Modular design with Active Cheerios}

The use of 3D printers allows us to design particles with different motion characteristics. Here, we demonstrate this by first designing chiral active Cheerios and then presenting a modular design where various particles are mechanically linked to program different propulsion trajectories.

We use knowledge of directional thrust in round particles with one or more outlets (see Section II and Fig.~\ref{fig:fig4}a and b) to create another new particle. The design achieves a chiral particle using two outlets with semicircular interfacial cups to diffuse fuel tangentially to the particle body circumference (see Fig.~\ref{fig:fig4}c). Redirecting the fuel deposition and direction of thrust creates a moment on the particle using outlet force principles and an angular rotation that depends on ethanol concentration (see Fig.~\ref{fig:fig4}d). Higher ethanol concentrations result in greater maximum angular velocity and slightly longer lifetimes. The particle start with high angular velocity, which gradually diminishes, following characteristically shorter lifetimes than single-outlet motors mainly because two outlets in the chiral particle cause it to release its fuel more quickly.

Combining the single-outlet and chiral particle, we built rigid, mechanically-linked multi-particle networks for customized locomotion (see Fig.~\ref{fig:fig4}e and f). The constructs are made by connecting individual particles with thin 3D-printed linkages that lock onto the fuel tank inlets located on the tops of the particles. We affix three particles to a straight linkage, allowing the overall assembly to float and move within a large square water-basin container. The linkage spacing is similar to the motor diameter, which minimizes interaction with the fast flows exiting the outlets of individual particles. Hence, we can program a trajectory based on the Marangoni-induced force vectors. By combining each motor propulsion mechanism, we construct diagnostic assemblies to confirm the modularity of our linked active particles.

Besides designing the particles (and turning them on/off by fueling or not), we also use their orientation to create a variety of movement modes (see Fig.~\ref{fig:fig4}f). The motions can be categorized as linear, curved, on-axis spinning, and off-axis spinning. We create a predominantly linear trajectory across the containment tank by powering both single-outlet particles attached to the construct and orienting them in the same thrust direction (Fig.~\ref{fig:fig4}f-i). During linear translation, lateral fluctuation is likely due to the structure's higher inertia and complex drag profile, creating symmetry breaking during transit. We create an assemblage with a curved trajectory using a linear configuration with a powered central chiral particle. When all particles are powered, the chiral one imparts a moment onto the assembly, producing a turning motion (Fig.~\ref{fig:fig4}f-(ii)). By reorienting one of the single-outlet particles by 180\textdegree, we create an on-axis spinning assembly, where the reinforcing moment from the single-outlet side particles in the direction of the chiral particle's counterclockwise rotation forms a chiral assembly that spins on its own axis (Fig.~\ref{fig:fig4}f-iii).  
We then demonstrate an off-axis spinning design using a single-outlet particle propelling parallel and outward from the constructed chain with an inactive chiral motor. The perpendicular Marangoni propulsion forces from the single-outlet designs produce off-axis spinning behavior (see Fig.~\ref{fig:fig4}f-iv).

These basic demonstrations validate that single Marangoni-driven active particle systems can be scaled to larger, more predictable modular assemblies by showing each unique propulsion behavior and trajectory. Such designer systems can be improved and further developed in various ways, as we will briefly discuss in the following conclusions and outlook section.

\section{\label{sec:level6}Conclusion \& Outlook}
This study demonstrates a straightforward yet effective approach to designing Marangoni-driven active particles that operate on a water surface, leveraging both the Marangoni and Cheerios effects to produce tunable motility and interaction patterns. 
The versatility of this system not only allows for precise control over particle behavior but also provides a simplified platform for studying complex collective dynamics. Such dynamics are analogous to those observed in natural systems, including ant rafts~\cite{ko2022small} and bacteria on surfaces~\cite{deng2020motile,panich2024swashing}, where interactions at interfaces give rise to emergent phenomena. By offering insights into the interplay of forces and interactions, this system has the potential to deepen our understanding of these biological systems and inform the development of bio-inspired designs in active matter and soft robotics. Moreover, the present system provides access to less-studied hydrodynamic regimes, including inertial effects~\cite{klotsa2019above}, making it a versatile tool for exploring new active matter behaviors. Besides ethanol concentration, one could also control the motility of the particles by the viscosity of the fuel or the liquid around it (see Appendix E). The experimental results also showcase other complex behaviors in multi-body systems, including phenomena such as active crystals~\cite{petroff2015fast,tan_odd_2022}, where a cluster of active Cheerios co-translate and co-rotate, and weak self-avoidance~\cite{daftari2022self,albers2024billiards}, where particles avoid previous trajectories at higher ethanol concentrations. Future studies could further explore the details of these features.

Designing active particles with various trajectories and also modular systems where particles of different characters are mechanically linked proves the flexibility of the present system. However, the design process for active Cheerios was primarily based on trial and error, but future work could benefit from integrating optimization techniques to refine the design pipeline for greater precision and efficiency but also more practical trajectories. Additionally, machine learning could help identify optimal geometries and fuel compositions more systematically, improving both motility and stability. Furthermore, programmable fuel release mechanisms could enable longer and more complex trajectories, opening up possibilities for increasingly sophisticated interfacial navigation.

Throughout this article, we primarily adopted an exploratory approach with minimal reliance on theory. The experimental results presented here could greatly benefit from future theoretical investigations across various scales. At the single-particle scale, the interplay of physicochemical hydrodynamics near the outlet, symmetry-breaking, and (inertial) propulsion offers a fertile ground for exploration. The simplicity of the experiments further enables precise testing of these theories. For multibody systems, experimental findings could be complemented by a range of theoretical frameworks, from simplified active particle models based on Langevin-like equations to detailed simulations accounting for hydrodynamic interactions and surface deformability.

Finally, we would like to mention that beyond research applications, the simplicity and adaptability of this active particle system make it highly suitable for educational purposes. It offers a hands-on opportunity to teach fundamental concepts in Marangoni flow, flow visualization, active matter, and particle tracking. With real-time visual feedback, students can engage deeply with the principles underlying these phenomena, making this system a valuable tool for both experimental science and science education. Overall, this work establishes a foundation for further exploration of active and adaptive interfacial devices with potential applications in robotics, environmental sensing, and beyond.


\section*{Acknowledgments}

We thank the Technology Centre of the University of Amsterdam for their assistance with the experimental setup. We also thank Vera Horjus, Corentin Coulais, and Vatsal Sanjay for valuable discussions. MJ acknowledges support from the NWO grant no. OCENW.XS21.1.140. JKW acknowledges the support of the National Science Foundation Graduate Research Fellowship Program for support during the writing and submission of this article.

\pagebreak
\setcounter{figure}{0}  
\renewcommand\thefigure{\thesection.\arabic{figure}}   
 
\section*{Appendices}
\section*{Appendix A: 3D Printing}
\label{sec:appA}
 
We print the particles using PLA (red Polylactic Acid Filament; Ultimaker, Utrecht, NL) fused-deposition modeling (Ultimaker 2+; Ultimaker, Utrecht, NL), due to its prevalence in extrusion-based printing for labs and classrooms. The containment ring is also 3D printed using PETG (Polyethylene terephthalate glycol) fused filament fabrication. We use the large 3D printed $45\,\mathrm{cm}$ diameter and  $3\,\mathrm{cm}$ deep ring to perform far-field motility in a  $6\,\mathrm{cm}$ deep rectangular water basin.

\begin{figure}
\centering
\includegraphics[width=320pt]{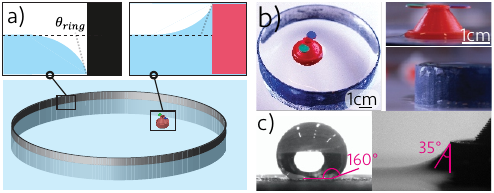}

\caption{Contact angle for the motors and containment ring in a water basin. a) The black wall shows the cross-sectional version of the containment ring with a hydrophobic surface, while the red wall is the cross-sectional surface of the particle and a hydrophilic surface. b) Photograph of the particle and containment. c) Approximation of the contact angle from side images of a water droplet on a coated hydrophobic surface and the contact angle of the meniscus on an active Cheerios.}
\label{fig:Sup1}
\end{figure}

\section*{\label{sec:AppB}Appendix B: Hydrophilic Particles and Hydrophobic Walls}

All of the printed materials are hydrophilic with a pronounced minescus with a contact angle of $\sim 35\,\mathrm{^\circ}$ on their surfaces when at the air-water interface (see Fig. V.1). We coat the containment ring with a super hydrophobic spray (NeverWet Multi-surface Spray; Rust-Oleum, Evanston, IL, U.S.). The hydrophobic ring has a contact angle of $\theta_{ring} \approx 160\,\mathrm{^\circ}$, which prolonged motility with reduced structure-structure interaction between the Active Cheerios and the ring.

\section*{\label{sec:AppC}Appendix C: Propulsion}

The theory of propulsion for active Cheerios deserves a more complete analysis. Here, we aim to present a simplified picture by balancing estimated forces. The surface tension gradient generated at the particle's outlet leads to a mismatch between direct forcing on the particle and, hence, propulsion. The magnitude of this force can be estimated as $\Delta \gamma D_n$, where $\Delta \gamma$ is the difference between the surface tension on two sides and $D_n$ is the nozzle diameter as a rough estimation of the length contact line force is acting on. The driving force should balance the drag force. The Reynolds number ($Re = \rho U_p D_p /\mu$, with $\rho$ as the density, $U_p$ as the particle velocity, and $\mu$ as the viscosity) is much larger than unity in most cases here. Hence, the drag force could be estimated as $\rho U_p^2 D_p^2$. This balance leads to $U_p \approx (\Delta \gamma \, D_n / \rho \, D_p^2) ^{1/2} \sim \mathcal{O}$(1 cm/s). On the other hand, the surface tension gradient at the outlet also leads to Marangoni flow. An estimation of fluid velocity at the nozzle may be achieved by balancing the Marangoni stresses $\Delta \gamma \, \mathcal{L}$, where $\mathcal{L}$ is a characteristic length scale where surface tension gradient is acting. Inspecting the flow field around the particles, we see a region of fast flow near the nozzle (see fig~\ref{fig:fig1}c). 
Assuming the gradient of surface tension is negligible beyond this region and based on observation, we take $\mathcal{L}\approx D_p$. 
If we assume the Marangoni stress at the nozzle balances with the viscous stress of a thin layer near the interface approximated as $\mu U_e / \mathcal{H}$, with $\mathcal{H}\approx D_n$, we then arrive at nozzle velocity of $U_e \approx D_n \Delta \gamma / D_p \mu$. 
Finally, balancing the flow forces generated at the nozzle with the drag force, we arrive at $U_p \approx U_e D_n / D_p \sim \mathcal{O}$( $1\,\mathrm{cm/s}$), which is the same order of magnitude as found from direct forcing on the contact line of the particle. Hence, in general, we expect both mechanisms be involved in particle's propulsion. However, note that, for most results shown here, we estimate Marangoni flow stresses play a more pronounced role.  
Better estimation of particle velocity, as well as particle trajectories, requires more complete theoretical and computational study of active Cheerios.

\section*{\label{sec:AppD}Appendix D: Inertial Effects}

The particles are inertial throughout their lifetimes. In single outlet particles, this result in a phase difference between particle orientation and velocity direction (See Fig. V.2). This is shown by comparing the angle of particle orientation (black) and velocity direction (red). When the particle decays, the difference between the two vectors become apparent as the propulsive effects are less effective, and the orientation of the particle is then static due to reduced fuel. 

\begin{figure}
\centering
\includegraphics[width=300pt]{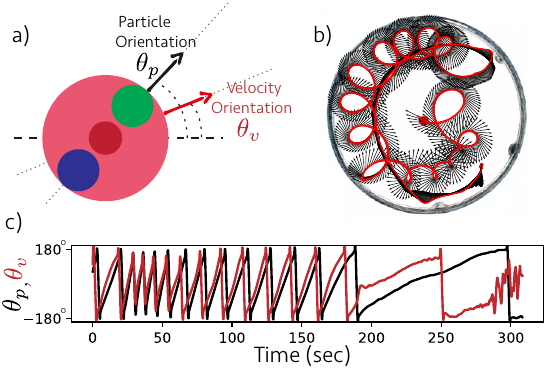}

\caption{a) Identification of the angles $\theta)p$ and $\theta)v$ for the actual particle orientation (black) based on the geometrical location of the fuel outlet and the tracked velocity orientation (red). b) Example of particle and velocity orientation vector arrows plotted during the lifetime of a particle with $50\,\mathrm{\%}$ ethanol fuel. c) Variation of $\theta_p$ and $\theta_v$.}
\label{fig:Sup2}
\end{figure}

\section*{\label{sec:AppE}Appendix E: Viscosity Effects}

Here, we report the results varying the fuel and reservoir liquid viscosity independently using glycerine. Glycerine is first added to the ethanol fuel to observe how the viscosity affects the particle lifetime, stabilizing and lowering the diffusion rate of the fuel  (see Fig. V.3a).hen using $50\,\mathrm{\%}$ glycerine - $50\,\mathrm{\%}$ ethanol, we get dramatically different qualitative and quantitative results. There is a clear velocity plateau at approximately $2.5\,\mathrm{cm/s}$ with a much more gradual decline to cessation. In this case, this shows that there are dramatically more consistent and prolonged lifetimes up to three times longer or death beginning at $750\,\mathrm{s}$. The lower ethanol also correlates to a lower overall average speed calculated for each particle. The velocity is dramatically lowered at lower ethanol percents, likely due to the severely impeded diffusion and more significant viscous dissipation. This informs us that future studies will investigate higher viscosity fuels to tune the inertial effects and achieve Stoksean swimmers if required. 

In a different experiment, glycerine is added to the reservoir basin to evaluate the effects of viscosity in the basin on lifetime and velocity (see Fig. V.3b). These basin liquid tests show that The highest concentration of pure glycerine corresponds to the lowest velocity of all tests conducted. In general, we observe increased velocity and lifetimes when reducing the glycerine concentration in the basin, showing the higher viscous stresses on the particles. The glycerine water basin was largely unsuccessful in extending lifetimes or generating more linear trajectory paths. The lower performance is likely due to the reduced fuel diffusion efficacy on the reservoir surface and less Marangoni propulsion. The fuel likely diffuses too rapidly around the motor with inadequate force transfer, lowering velocity and acceleration. 

\begin{figure}
\centering
\includegraphics[width=300pt]{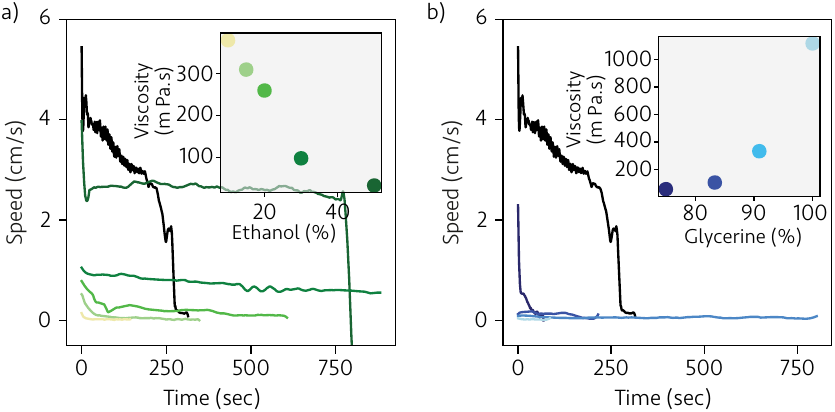}

\caption{a) Traveling speed of particles with glycerine-ethanol fuel. The viscosities (inset) are measured with an Anton Paar 502 rheometer. b) Traveling speed of particles with ethanol solution as fuel, moving in a basin with water-glycerine solutions. }
 \label{fig:cheerios_glycerine} 
\end{figure}

\newpage
\printbibliography


\end{refsection}

\clearpage
\setcounter{page}{1}
\renewcommand{\thefigure}{S\arabic{figure}}
\setcounter{figure}{0}
\renewcommand{\theequation}{S\arabic{equation}}
\setcounter{equation}{0}


\end{document}